**Evaluating the Impact of Automated Vehicles on Residential Location Distribution using Activity-based Accessibility: A Case Study of Japanese Regional Areas**

**Lichen Luo (corresponding author)**
Department of Urban Engineering
The University of Tokyo, Tokyo, Japan, 113-8656
Email: lichen.l@ut.t.u-tokyo.ac.jp

**Giancarlos Parady**
Department of Urban Engineering
The University of Tokyo, Tokyo, Japan, 113-8656
Email: gtroncoso@ut.t.u-tokyo.ac.jp

**Kiyoshi Takami**
Department of Urban Engineering
The University of Tokyo, Tokyo, Japan, 113-8656
Email: takami@ut.t.u-tokyo.ac.jp

Word Count: 6,231 words + 5 tables (250 words per table) = 7,481 words

Submitted Date: August 1st, 2022

**ABSTRACT**

Automated Vehicles (AVs) are expected to disrupt the transport sector in the future. Extensive research efforts have been dedicated to studying its potential implications. However, the existing literature is yet limited regarding the long-term impacts. To fill this gap, this paper estimates and validates a residential location choice model to evaluate the impacts of AVs on residential location distributions in a context of Japanese regional area. Activity-based accessibility is used to reflect the changes from AVs in transport costs. The year 2040 is set as the backdrop for the analyses, where the effects of the decreased population are reflected in the scenario settings, along with some other variables to accommodate the uncertainties in the characteristics of AVs. The simulation results confirm the potential of urban expansion. The results demonstrate that, compared to Base Scenario, the median distances between the residences and the closest Dwelling Attraction Areas expand by 7.2% and 41.6% for two AV scenarios, respectively. Two hypothetical policy mandates are then applied to alleviate the problem. The results suggest that providing a 20% subsidy to the land price is effective for the scenario with relatively conservative AV settings, as the median distance indicator can be resumed to the level of Base Scenario.

**INTRODUCTION**

Automated vehicles (AVs) are currently expected to be a promising mode of travel in the future given many original features differentiated them from Human-driven vehicles (HVs). These features including the less general cost of travel are mostly derived from no driver intervention required in high automation (*1*; SAE Level 4 or 5) anymore.

Despite that extensive research efforts are ongoing to study AV features and better understand their potential effects, implications on residential location choice are still understudied (*2*). Two opposite AV effects on residential locations are usually assumed (*3*). First, AVs could fuel re-urbanization given the reduced need to find parking lots in particularly costly urban areas. Second, AVs could fuel urban sprawl since the driverless feature could make long trips to the higher-density areas less burdensome. Furthermore, despite some previous study (*4 错误!未找到引用源。*) revealed that different level of benefits could be enjoyed between the urban and suburban area, the studies on AV impacts on residential location in the context of Japan is yet missing.

Against these backgrounds, this article intends to offer insights into quantifying the privately-owned AV (PAV) implications on residential location distribution focusing on a Japanese regional area. We particularly consider the second effect mentioned above.

Such an investigation necessitates building models with the properties to reflect transport changes in long-term land use models. For that purpose, this study uses activity-based accessibility (*5*) that generated from a short-term transport simulation model system (*6*) as the connection. A multinomial logit (MNL)-based residential location model is estimated, validated, and simulated with the PAV effects reflected in the scenario settings. Two policy mandates are then tested to attempt to mitigate the possible negative effects.

The rest of this paper is organized as follows: the next section reviews existing literature evaluating AV implications on residential choice; then a section describing the methodology and data sources follows; modeling results are presented in the subsequent section; finally, the last section states the conclusion and discussion.

**LITERATURE REVIEW**

Using survey methods is the first common stream of research in studying AV impacts on residential location choice. Some early survey studies did not quantify the magnitude of the changes but the willingness to shift residence. For example, Bansal et al. 2016 (*7*) found that 74% of their Austin respondents believed no home location change would be made following the AV introduction, while 14% and 12% chose to move closer and farther from the city centers, respectively. Kim et al. 2020 (*8*) also found the majority (77.3%) of the respondents in Georgia expected no change in residential locations. Nonetheless, the dual effects were revealed: the more people expected AVs to make their time using more productively, the more likely they would move away from work and other currently frequently-visited places; it was, however, the opposite for those with fewer vehicles.

Some other survey studies quantified the magnitude of the changes in residential locations. Moore et al. 2020 (*9*) predicted a 68% increase in commute time conditional on relocation based on their Dallas-Fort-Worth metropolitan area-based survey. Hence, they argued it suggested the upper range of potential urban sprawl.

More commonly used to quantify AV residential choices were simulation studies. To the authors' knowledge, there are only four works belonging to this category, they are introduced below with relatively more details.

Gelauff et al. 2019 (*10*) studied the effects of PAVs on residential location change with a model designed in land use transport integrated (LUTI) fashion to connect transport and land use effects. Upon the change in the housing demand, land prices were adjusted until the housing demand and supply equaled again. Three models: a commuting mode model, a job location model, and a residential location model were combined within a nested logit (NL) framework. Under assumptions such as a 20% decrease in the travel time coefficient of privately-owned cars (for trips over 5km), home-job location distance would increase by 16.8% across the Netherlands.

Meng et al. 2019 (*11*) employed a LUTI simulation platform SimMobility to study the Shared AV (SAV) impacts on moving patterns in Singapore. The study combined a long-term housing market model, a job location choice model, and a household vehicle ownership model, where the effects from SAV were reflected as logsums calculated from the transport models proposed in their previous studies. Housing bidding behaviors were explicitly considered. Their results suggested a roughly slight moving out tendency from the central region under a scenario with SAV added competing with the existing transport modes. Nevertheless, under a scenario where only SAV and public transits were allowed, 15.6% more people moving into the central region were found.

Zhang and Guhathakurta, 2021 (*12*) also focused on residential location choice in exclusively "the era of SAV". The authors connected an MNL-based residential location model and an agent-based SAV simulation model built in their previous study. The simulation model took the trips generated through a Four-step model as the input for Atlanta, Georgia. Scenarios were assumed where all trips would be served by SAVs whose in-vehicle travel time decreased by 25% to 100%. The general travel cost change was passed to the residential location model. The results found that some people would move away from their job locations, but some choose to move closer to the central areas. For example, people younger than forty without kids would relocate away from the city centers (6.8% median distance increase) and their job locations (23.4% median distance increase), while older people with children would move closer to the city centers (6.5% median distance decrease) but still away from their job location (20.9% median distance decrease).

Llorca et al., 2022 (*3*) studied the potential PAV effects on residential location choice in the context of Munich, Germany. They proposed a combination of a land-use model system, a travel demand model, and a traffic assignment model, as their LUTI methods. They tested eight AV scenarios covering the effects of changes in the value of travel time, parking restrictions, and traffic congestion. The commuting time changes largely account for the AV impacts connecting the transport models to the land use model. Their results observed urban sprawls. For example, for the full scenario where all the AV effects were included, those who work and also live in the city centers would be 2-3% fewer than in the non-AV scenario.

In summary, AV impacts on residential locations have not received enough research attention to date. The few studies provided results in conflict that "stems from the limited integration between simulation and behavioral studies" (*2*). Despite LUTI models having been widely applied, appropriately reflecting the changes from AVs in the residential location model is far from reaching a common ground. The results also varied widely depending on the contexts of study regions. Evaluating the AV impacts in the context where the related study has been missing should offer insights regarding this unsettled research question.

## METHODS

A multi-hierarchical LUTI model system is adopted as the methodology framework of this study (**Figure 1**). This model system mainly consists of three models: an activity-based travel demand model, DAS (*13*); an agent-based dynamic traffic simulation model, MATSim (*14*); and an MNL-based residential location model. As Figure 1 shows, the DAS models the travel demand, and MATSim models the travel supply component. These two models are integrated to capture the short-term travel demand-supply interactions: the outputs from both models, respectively time-specific OD pairs data and network conditions data, are exchanged across the two models. Exercising travel demand-supply equilibrium is necessary in, for example, measuring indirect effects such as traffic congestion due to induced travel from AVs. After the two models converge to a demand-supply equilibrium, activity-based accessibility (ABA; *5*) is calculated as a composite change from the transportation system to be passed to the residential location model as one of its inputs. The approach to incorporating the ABA follows Ben-Akiva and Bowman, 1998 (*15*). The texts in the grey background: job location and facility development are considered exogenous, i.e., either unchanged or set by scenarios. The details of building and implementing the short-term simulators are not the focus of this paper, readers are referred to Luo et al. 2022 (*6*).

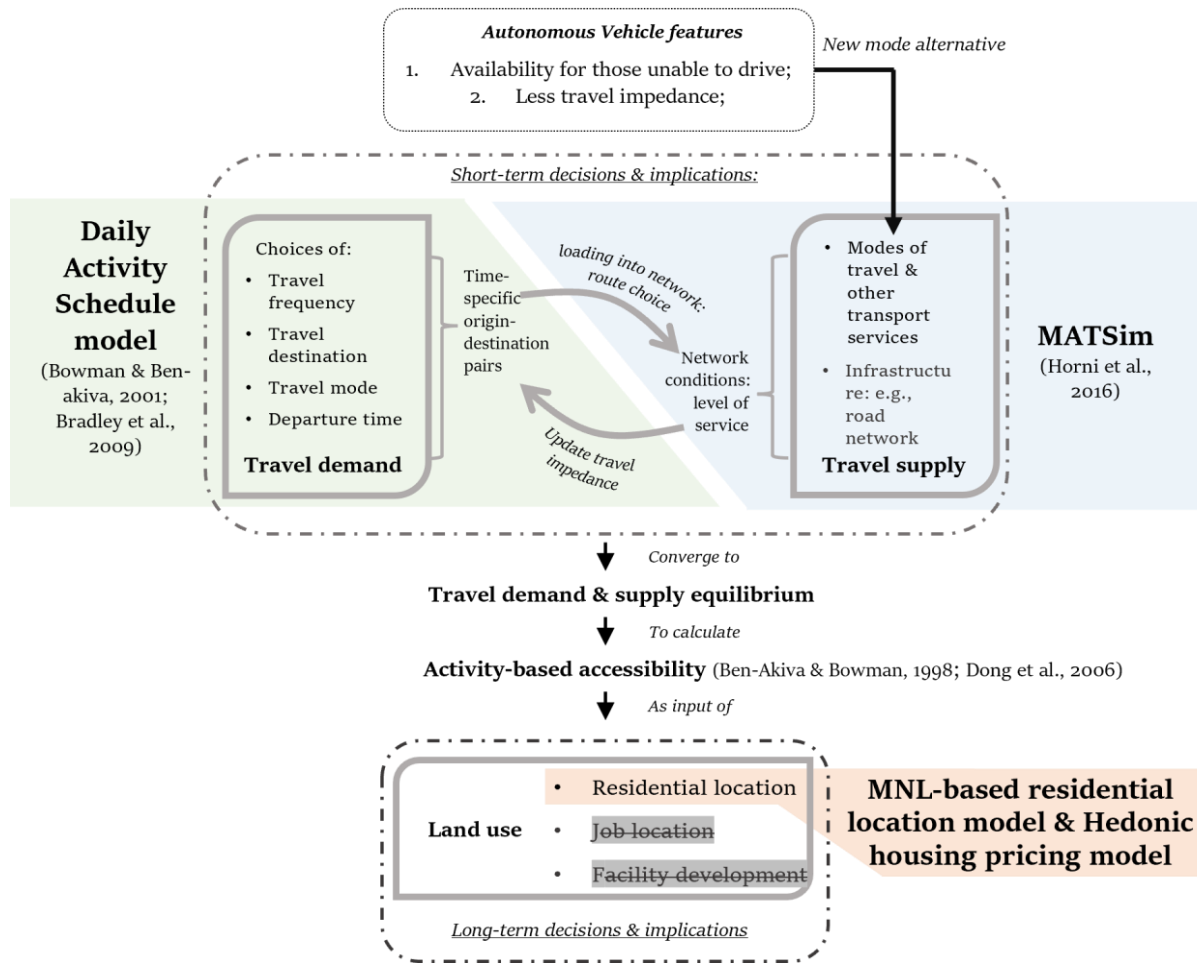

**Figure 1 General Methodology Framework**

**Study Region**

This article uses Gunma Prefecture of Japan as the study region to study potential AV implications. Gunma is a landlocked prefecture in the Capital Region, with its prefectural government located approximately 100km away from Tokyo Station (**Figure 2**).

Gunma covers 6,363 km$^2$ and has a total population of 1,940,333 as of 2020 (*16*), with a population density of 304.9 people per km$^2$. This value is slightly less than the average population density in Japan of 338.4 people per km$^2$ (*16*). Gunma is commonly known as a car-dependent society. The prefecture has the highest average private four-wheeled vehicle ownership in Japan with 0.705 vehicles per capita (*17*).

**Data Sources**

*Initial Travel Demand and Residential Location Data*

The Gunma 2015 Person Trip Survey data (PT data) is used as the initial travel demand and Residential Location Data. The government-funded survey was conducted by distributing questionnaires to 242,000 randomly sampled households in Gunma prefecture plus Ashikaga City in Tochigi prefecture across 2015 and 2016. The questionnaires collected demographic and socioeconomic information from each household, and travel records for one weekday. Note that some valuable information was not collected including individual income, financial and time budget.

The Gunma PT Data was pre-processed to filter missing or inexplicable attributes. Finally, a dataset of 16,425 households with 33,300 persons is adopted as the effective initial travel demand and residential location data. The sample constitutes 1.57% of the whole target population in 2015 (*18*). The dataset was randomly split into two 80% and 20% parts at the household level for the subsequent estimations and validations.

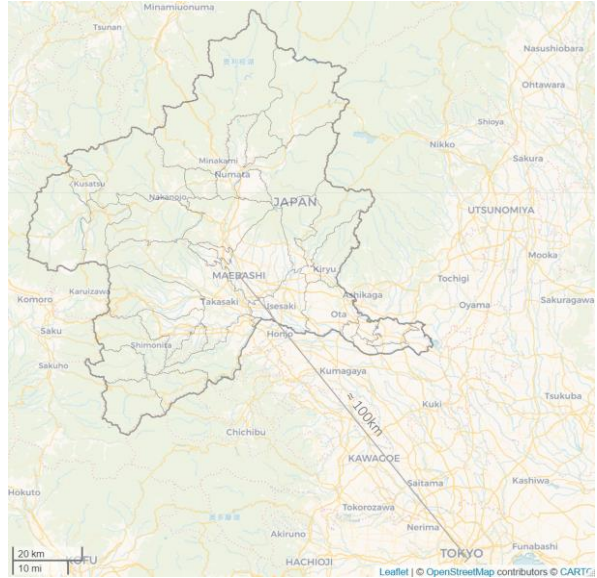

**Figure 2 Location of Gunma Prefecture. Source: OpenStreetMap**

*Land Use Data*

The analysis spatial resolution of this study is the $1km^2$ mesh cell. Such a level is considered appropriate as a balance between the requirement of spatial resolution and computational cost. Mesh-cell-level land use data were collected mainly from the Regional Mesh Statistics (*19*). After processing, a dataset with 3,001 mesh cells is obtained and will be used in this study.

Two important concepts of land use: Urban Function Attraction Area (UFAA) and Dwelling Attraction Area (DAA) that will be used in evaluations are introduced here. These concepts from Location Normalization Plan (*20*) refer to the target areas to attract respectively urban functional facilities (e.g., commercial, educational, and medical facilities) and residents in achieving compact urban structures. They are used to define the urban center areas in this study. The spatial distributions of UFAA and DAA are shown in Figure 3.

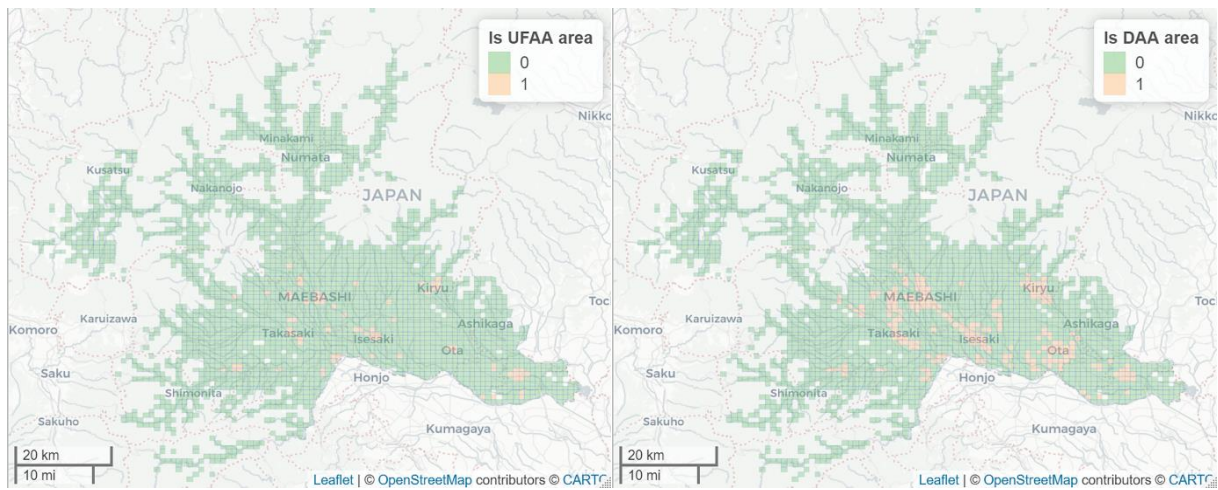

**Figure 3 (Left) Distribution of UFAA in Gunma PT Data Area, (Right) Distribution of DAA in Gunma PT Data Area. Source: adapted from the Municipal governments' websites**

Some other necessary attributes in modeling residential locations, however, are currently not available in Japan for the same resolution used in this study. Among others, the number of housing stocks and housing price per mesh cell are important as they measure from the supply side in the residence market.

For the housing stock data, we allocated the city-level data (*21*) to each mesh cell with some mesh-cell-level attributes such as land use type area ratio of each mesh cell (*22*) as the weights to approximate the mesh-cell-level data.

We used Land Value Publication Data (*23*) and a hedonic model to approximate the mesh-cell-level housing price data. The land price will be used as a proxy for housing price. In total, land prices of 824 dwellings located inside the Gunma PT data area were obtained. The dwelling-level data were spatially merged to 557 mesh cells. Then, we estimated a hedonic model for land price which took the mesh-cell-level land price values as the observed data. The details of the hedonic model are introduced next.

**Hedonic Models for Land Price**

The hedonic estimation results are shown in **Table 1**. For simplicity, spatial issues such as spatial dependence and spatial heterogeneity are not considered for this specification. Three logsum variables are added as independent variables in the Hedonic Model. They are calculated from the Tour Mode and Destination Level of the DAS model (*6*) to represent the tour-based transport accessibilities for each mesh cell.

**TABLE 1 Estimation Results of Land Price Hedonic Model**

| *Dependent variable: natural log of land price (JPY/m²)* | | |
|---|---|---|
| **Variable** | **Coefficient** | **T Value** |
| Intercept | 7.83 | *27.52* |
| #Housing stock | | *-0.64* |
| Tour-based logsum of work-purpose | 0.15 | *5.02* |
| Tour-based logsum of education-purpose | 0.043 | *6.91* |
| Tour-based logsum of other-purpose | 0.13 | *4.91* |
| Is Takasaki City | 0.30 | *6.37* |
| Is Maebashi City | 0.21 | *4.37* |
| Is Ota City | | *-0.34* |
| Is Isesaki City | | *-0.89* |
| Is Kiryu City | -0.11 | *-1.81* |
| Share of agricultural use area | -0.48 | *-3.80* |
| Share of forest area | | *-0.46* |
| Share of freshwater use area | | *-0.15* |
| Share of industrial use area | -0.58 | *-2.27* |
| **#Count** | 557 | |
| **Adjusted R squared** | 0.728 | |
| **F statistic** | 115.4 | |

The results indicate that mesh cells being transport accessible for all three types of tour, and with less area covered for either agricultural or industrial use are likely to have high land prices. Being in big cities such as Takasaki or Maebashi also has extra value on the land price. The high R squared suggests the generally good performance of the model.

The predicted land price as the result of the hedonic model is shown in **Figure 4**.

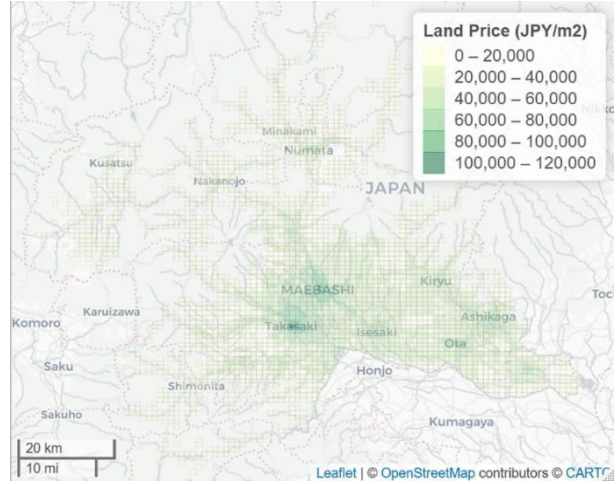

**Figure 4 Predicted Land Price by Mesh Cell**

**Residential Location Model Specification and Estimation**

This research follows Ben-Akiva and Bowman, 1998 (*15*) to integrate the residential location model with the activity-based model system (*6*). The residential model is specified as an MNL model at the household level, where the observed component of the utility of residential location $l$ for household $h$ is:

$$V_{h,l} = \beta X_l + \frac{\alpha_w}{W_h} \sum_{w \in W_h} A_{w,h,l} + \frac{\alpha_s}{S_h} \sum_{s \in S_h} A_{s,h,l} + \frac{\alpha_u}{U_h} \sum_{u \in U_h} A_{u,h,l}$$

Where $\alpha$, $\beta$ are coefficients to be estimated; $X_l$ is the attributes of location $l$; $A_{w,h,l}$, $A_{s,h,l}$, $A_{u,h,l}$ are ABA values (*5*) of worker, student, and unemployed people in household $h$ given residential location $l$, respectively; and $W_h$, $S_h$, $U_h$ are the number of each category in household $h$. Specifically, $A_{i,h,l}$ of individual $i$ is:

$$A_{i,h,l} = \ln \left( \sum_{p \in P_i} e^{V_p} \right) + Constant$$

Where $p$ is the daily activity pattern in the daily activity pattern choice set $P_i$ of individual $i$, $V_p$ is the systematic component of the utility of $p$.

These ABA terms are pre-calculated and estimated as independent variables with corresponding coefficient $\alpha$. The ABA term of one person category would not be included if that household has no member of that specific person category.

The ABA terms were normalized before the estimation to be comparable among different individuals. Specific procedures are shown below (*5, 24*):

$$A_i^{normalized} = \frac{A_i - A_i^{original}}{s_{i,t}}$$

Where $s_{i,t}$ is the scaling factor that approximates the marginal utility of travel time $t$. It was calculated as the change in ABA with 1 unit change, i.e., 1min, of $t$.

$$s_{i,t} = \frac{A_i^{\Delta t} - A_i}{\Delta t}$$

And $A_i^{original}$ is the original ABA before the change in policy or the transport system, which in this case refers to the ABA given the current home location of $i$'s household.

For each household, 50 mesh cells were sampled from the whole mesh cell dataset with Importance Sampling with Replacement (*25*) as the choice set of the model. Land price was used as the sampling weight in this procedure. Two correction terms concerning aggregated alternatives and alternative samplings are added (*25*): the number of housing stocks by each mesh cell was used as the size variable; the natural log of the inverse of the sampling probability was used to cancel out the bias in the alternative sampling.

We used the randomly sampled 80% estimation dataset of Gunma PT data for the residential model estimation. 13,140 households are exogenously divided into five market segments (*12*) of similar size by the age of the household head and number of household members to accommodate the demographical heterogeneity. The estimation results of the residential location model are shown in **Table 2**.

**TABLE 2 Estimation Results of Residential Location Model**

| Segments | Segment #1 | | Segment #2 | | Segment #3 | | Segment #4 | | Segment #5 | |
|---|---|---|---|---|---|---|---|---|---|---|
| **Age of the head of household** | (6,50] | | (6,50] | | (50,100] | | (50,65) | | [65,100] | |
| **#Household members[1]** | 3 or more | | 1 or 2 | | 3 or more | | 1 or 2 | | 1 or 2 | |
| **Variable** | Coef. | T Val. | Coef. | T Val. | Coef. | T Val. | Coef. | T Val. | Coef. | T Val. |
| Household average ABA for workers | 0.63 | *57.86* | 0.62 | *57.91* | 0.40 | *45.24* | 0.43 | *45.93* | 0.33 | *34.65* |
| Household average ABA for students | 0.060 | *8.35* | 0.12 | *6.36* | 0.026 | *2.47* | | *0.60* | | *0.98* |
| Household average ABA for unemployed people | 0.17 | *10.45* | 0.28 | *11.43* | 0.24 | *19.18* | 0.37 | *23.56* | 0.61 | *49.91* |
| Land price (10,000 JPY/m²) | -1.24 | *-21.81* | -0.74 | *-13.89* | -0.84 | *-17.05* | -0.82 | *-15.57* | -0.73 | *-16.73* |
| Share of building use area | -1.17 | *-10.89* | -1.21 | *-11.84* | -0.56 | *-5.67* | -0.75 | *-7.34* | -0.58 | *-6.90* |
| Share of agricultural use area | -2.09 | *-10.26* | -1.77 | *-8.45* | -0.92 | *-4.99* | -1.25 | *-6.12* | -0.85 | *-5.04* |
| Share of freshwater area | | *-0.06* | | *1.15* | 0.31 | *1.91* | | *0.98* | 0.41 | *2.66* |
| Share of forest area | 0.74 | *3.34* | 0.70 | *3.13* | 1.18 | *6.43* | 1.16 | *5.78* | 1.41 | *8.76* |
| Is Takasaki city | 0.73 | *5.86* | | *0.038* | 0.42 | *3.91* | 0.30 | *2.51* | 0.64 | *6.57* |
| Is Maebashi city | 0.24 | *2.28* | | *-0.42* | 0.26 | *2.93* | 0.29 | *2.98* | 0.73 | *9.14* |
| Is Ota city | -0.73 | *-7.97* | -0.59 | *-6.67* | -0.48 | *-5.90* | -0.40 | *-4.58* | 0.13 | *1.82* |
| Is Isesaki city | -0.69 | *-7.39* | -0.61 | *-6.60* | -0.54 | *-6.26* | -0.52 | *-5.61* | | *-0.94* |
| Is Kiryu city | -0.37 | *-3.15* | -0.32 | *-2.67* | -0.24 | *-2.54* | -0.35 | *-3.21* | | *-1.56* |
| #Employees of primary and secondary sector | -0.10 | *-13.31* | -0.054 | *-9.30* | -0.065 | *-8.86* | -0.067 | *-8.70* | -0.023 | *-4.55* |
| #Employees of tertiary sector | -0.033 | *-9.08* | -0.038 | *-11.87* | -0.047 | *-11.42* | -0.022 | *-6.97* | -0.048 | *-47.82* |
| Size variable: #housing stock | 1.00 | - | 1.00 | - | 1.00 | - | 1.00 | - | 1.00 | - |
| **#Observations** | 2,630 | | 2,554 | | 2,578 | | 2,192 | | 3,186 | |
| **Initial likelihood** | -9,477.81 | | -9,128.47 | | -9,407.29 | | -7,953.38 | | -11,534.37 | |
| **Final likelihood** | -6,406.43 | | -6,142.27 | | -7,586.59 | | -6,335.93 | | -9,561.44 | |
| **Adjusted rho squared** | 0.322 | | 0.325 | | 0.192 | | 0.201 | | 0.170 | |

---

[1] *Household members with age less than six were counted, but they were not considered in subsequent ABA calculations.*

The results show expected coefficient signs of household average ABA and land price, which indicate that the trade-off between the transportation and housing cost is captured. Besides that, all households are found to prefer mesh cells with fewer buildings, fewer farmlands, more forests, fewer employees no matter the job category, and governed by the two big cities, Takasaki and Maebashi, all else being equal. The preferences vary across different market segments: for example, those small-size households with senior heads (Segment #4 and #5) care not about transport accessibilities for students in their residential location choices. This makes sense as there usually is no student in these two types of households.

**Residential Location Model Validation**

We conducted internal validation (*26*) for the residential location model to prove its reproducibility and reliability. The 20% validation dataset of Gunma PT data was used. Validation results were averaged from 10-time Monte-Carlo simulations of the residential location choice model to mitigate the random sampling error.

We chose two indicators for the comparisons between the observed results and simulated results from the validation dataset. The first type of indicator is network distance to the closest center area: UFAA and DAA, to represent the feature of the residence pattern in a polycentric area like Gunma. The second indicator is the count of households residing in DAA, which we consider is more straightforward to serve as a policy evaluator. **Table 3** shows the validation results regarding the indicators.

**TABLE 3 Residential Location Model Validation Results**

| | | Mean | Median | Min. | Max. | Standard Deviation |
|---|---|---|---|---|---|---|
| **Network distance (m) from residences to the closest DAA** | Observed results | 3,200 | 1,285 | 0 | 62,796 | 7,147 |
| | Simulated results | 2,324 | 1,273 | 0 | 63,726 | 5,373 |
| | Observed results (removing data farther than 10,000m) | 1,732 | 1,225 | 0 | 9,974 | 2,210 |
| | Simulated results (removing data farther than 10,000m) | 1,632 | 1,245 | 0 | 9,805 | 2,008 |
| **Network distance (m) from residences to the closest UFAA** | Observed results | 4,438 | 2,435 | 0 | 64,385 | 7,064 |
| | Simulated results | 3,644 | 2,340 | 0 | 65,493 | 5,534 |
| | Observed results (removing data farther than 10,000m) | 2,863 | 2,203 | 0 | 9,972 | 2,364 |
| | Simulated results (removing data farther than 10,000m) | 2,720 | 2,153 | 0 | 9,852 | 2,256 |
| **#Household (shares) residing in DAA** | Observed results | 1,328 (*40.4%*) | | | | |
| | Simulated results | 1,333 (*40.6%*) | | | | |

According to the comparisons of the summary statistics in **Table 3**, the estimated residential location choice model has good reliability in predicting the median distance to both the nearest UFAA and DAA and the number of households residing in DAA. This is not the case, however, for mean values where the simulated value is around 20% smaller than the observed one. The discrepancies are well explained in **Figure 5** and **Figure 6** which illustrate the spatial distributions. Extreme values affect greatly the mean values, as the shares of those who reside more than 10 or 15km away from their closest UFAA and DAA are underpredicted. Some variables concerning the historical influence (e.g., some seniors just prefer to keep living where their ancestors have long been residing) could be added in the future model estimation, as we are currently not accessible to such data. In fact, average values after removing those farther than 10km are much closer between the observed and simulated results (**Table 3**).

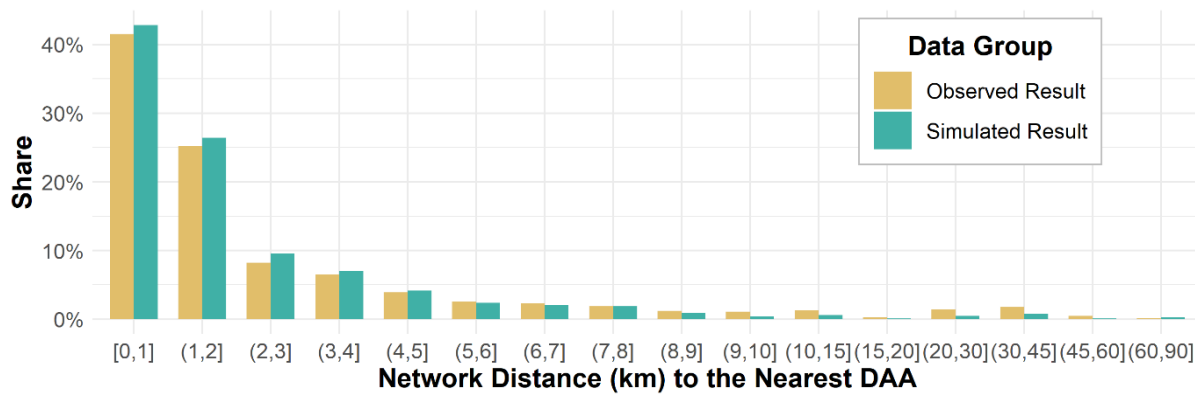

**Figure 5 Validation Results of Distribution of Residences' Distance to the Nearest DAA**

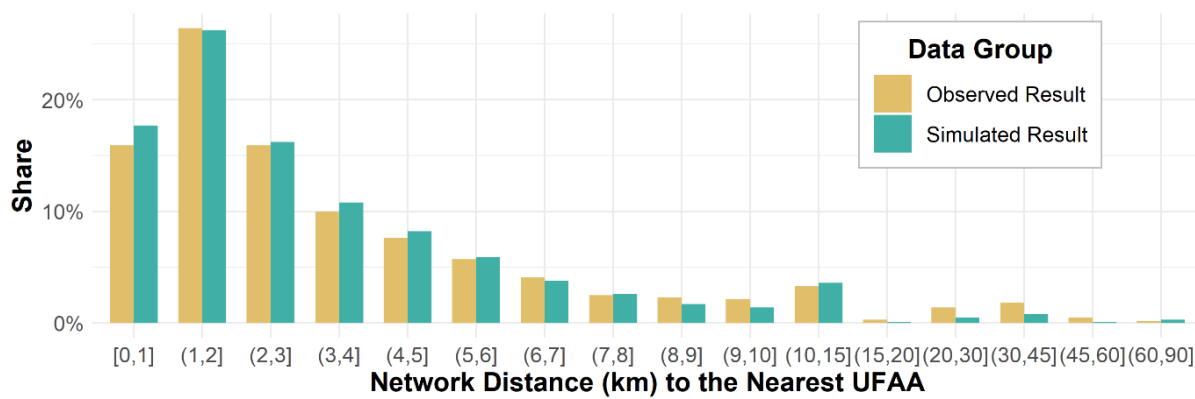

**Figure 6 Validation Results of Distribution of Residences' Distance to the Nearest UFAA**

In summary, the estimated model shows sufficient reproducibility and is adequate for forecasting. However, as bias caused by the extreme values is confirmed, the subsequent analyses will focus on the median statistic for the distance indicators.

## MODEL APPLICATION AND EVALUATIONS

In this section, we describe the model settings and simulation results of how residential location patterns would change under the AV scenarios. Furthermore, the results of two policy mandates as attempts to mitigate the potential side effects are also presented.

### Model Application Settings

As a common approach to reflect potential variations in AV characteristics, scenario analysis was used in this study. The forecast year for the scenario analysis is 2040, a time point assumed for the prevalence of AVs (*27*). A full PAV ownership penetration level is assumed, that is, households that currently own HVs are assumed to switch to PAVs, and all the members in these households are assumed to have access to a PAV irrespective of their driver's license ownership status. A ratio of 82.45% to the current population is assumed for 2040 based on the predictions made by NIPSSR, 2018 (*28*). The demographic patterns across the population are assumed the same as in 2015.

A base scenario without AVs but only population changes is firstly included. Two AV scenarios are then defined with two levels of value of travel time (VoT) and one level of road capacity improvement assumed: in Scenario 1, VoT of AVs is set 75% and 85% of HVs for commuting-purpose tours and other-purpose tours, respectively; in Scenario 2, VoT of AVs is set 50% and 70% of HVs for commuting-purpose tours and other-purpose tours, respectively; AVs would gain 20% improvement in road capacity compared to HVs in both scenarios. Reasonings for these settings are referred to the previous work (*6*).

Regarding the residence moving-or-not choice, we assumed that the households would make residence moving choices following the observed moving choice results currently in the study region. To elaborate, the moving choice is applied based on the data from the National Census (*18*) to identify the probability of whether households still reside where they were 5 years ago. The probability of whether a household would move by 2040 is calculated as one minus the fifth power of the did-not-move ratio between 2010 and 2015. Fifth power means that there are five periods of five years between 2015 and 2040. The ratios were differentiated by the age of the head of the household. The calculated moving probabilities vary with the age of the head of the household, ranging from 16.1% for the age of the head of household over 85 to 100.0% for the age of the head of household between 15 and 29. Complete results are not included for the sake of conciseness. Monte-Carlo simulations will be run to decide whether the household would move before simulating the residential location choice.

For the residential location choice simulation, the treatments are similar to what was applied in the model validation section. 50 alternatives of mesh cells sampled with Importance Sampling with Replacement (*15*) were provided to each household as the choice set. ABA values were pre-calculated for each sampled alternative given the AV settings.

## Simulation Results

The residential location model was simulated 10 times in a Monte-Carlo fashion, and the results were averaged the same as in the model validation section. The short-term simulation results including ABA were presented in Luo et al. 2022 (*6*).

**Figure 7** shows the spatial distributions of the residences by mesh cell for Base Scenario. **Figure 8** shows the results of the AV scenarios as the difference values against Base Scenario. The same legends apply to the two AV scenarios for direct comparisons.

Compared to the Base Scenario, a clear moving trend away from the central cities is found in both scenarios, while Scenario 2 is with higher extents. The moving trend can be better identified by the summary results of the evaluators shown in **Table 4**.

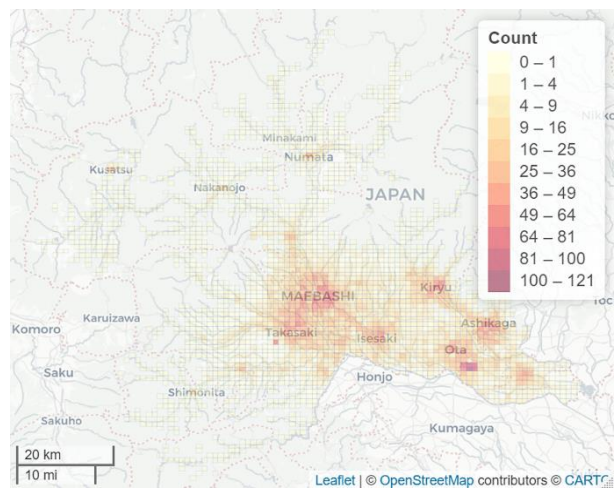

**Figure 7 Number of Residences by Mesh Cell: Base Scenario**

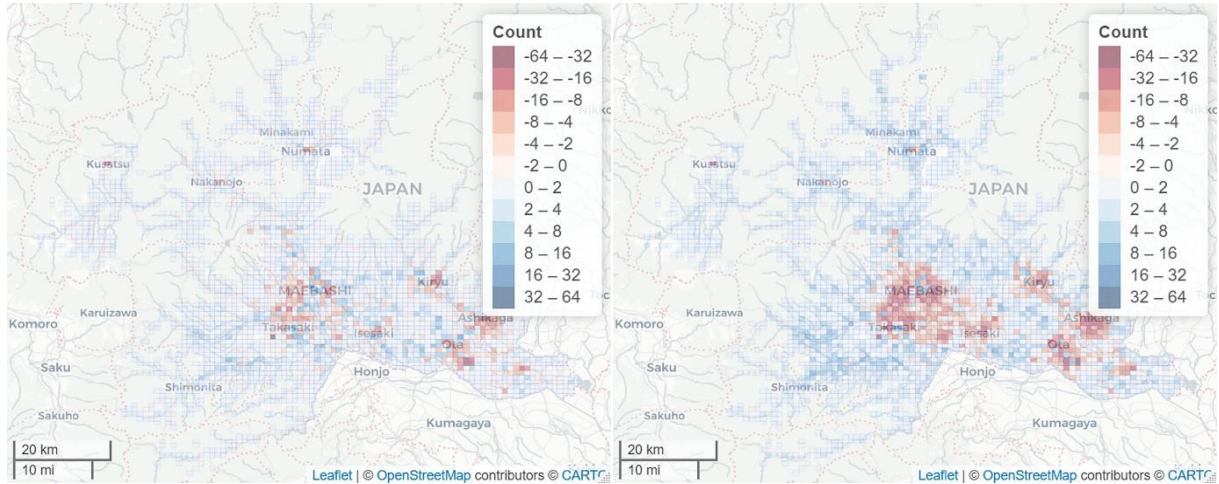

**Figure 8 Number of Residences by Mesh Cell: (Left) Scenario 1 against Base Scenario, (Right) Scenario 2 against Base Scenario**

**TABLE 4 Simulation Results Summary**

| Scenarios | PT Data | Base Scenario | | Scenario 1 | | Scenario 2 | |
|---|---|---|---|---|---|---|---|
| ***#Household*** | *16,425* | *13,542* | | | | | |
| **Measures** | Value | Value | Percentage change against PT | Value | Percentage change against Base | Value | Percentage change against Base |
| **Median network distance from residences to the closest DAA (m)** | 1,290 | 1,405 | +8.9% | 1,506 | +7.2% | 1,990 | +41.6% |
| **Median network distance from residences to the closest UFAA (m)** | 2,549 | 2,731 | +7.1% | 2,803 | +2.6% | 3,538 | +29.5% |
| **Share of Households Residing in DAA** | 40.2% | 36.2% | -10.0% | 33.2% | -8.3% | 27.3% | -24.6% |

We can again identify clear moving trends away from both DAA and UFAA. Even for the Base Scenario that differs from the PT data only in population size, the median distance to the residents' nearest DAA and UFAA increase by 8.9% and 7.1%, respectively. These are presumably due to the improved level of service in the road network. With the introduction of PAVs, these two values escalate to at most (under Scenario 2) 41.6% and 29.5%, respectively. While for scenario 1, much more moderate increases are observed. The increases suggest the level of re-balance in the housing price and transport cost tradeoff: the residents are attracted by the reductions in travel impedance to live farther from the city centers.

As for the share of residents in DAA, 33.2% and 27.3% are reported in Scenario 1 and Scenario 2. The values decrease by 8.3% and 24.6% compared to the Base Scenario, respectively. The magnitude of changes is consistent with the distance indicator, where Scenario 1 shows moderate changes.

**Figure 9** and **Figure 10** show the shares of residence by distance bins to the nearest DAA and UFAA, respectively. These spatial distribution results generally confirm the findings from **Table 4**.

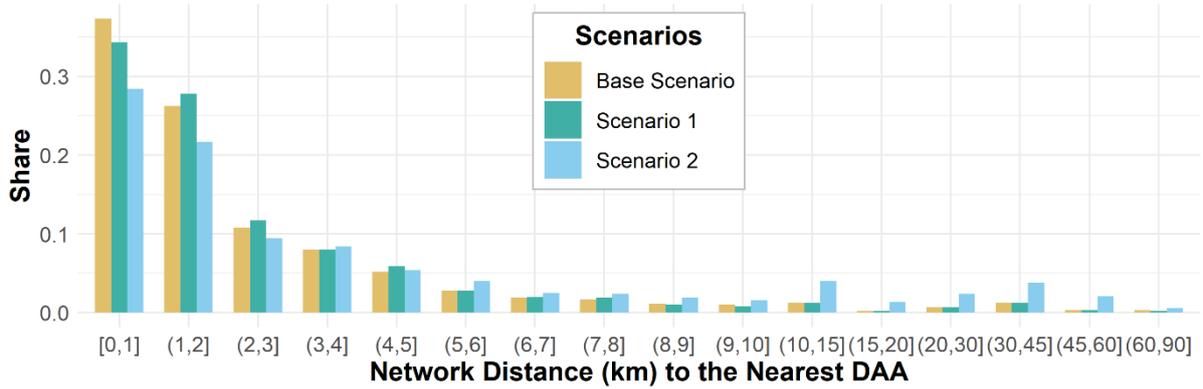

**Figure 9 Distribution of Residences' Distance to the Nearest DAA**

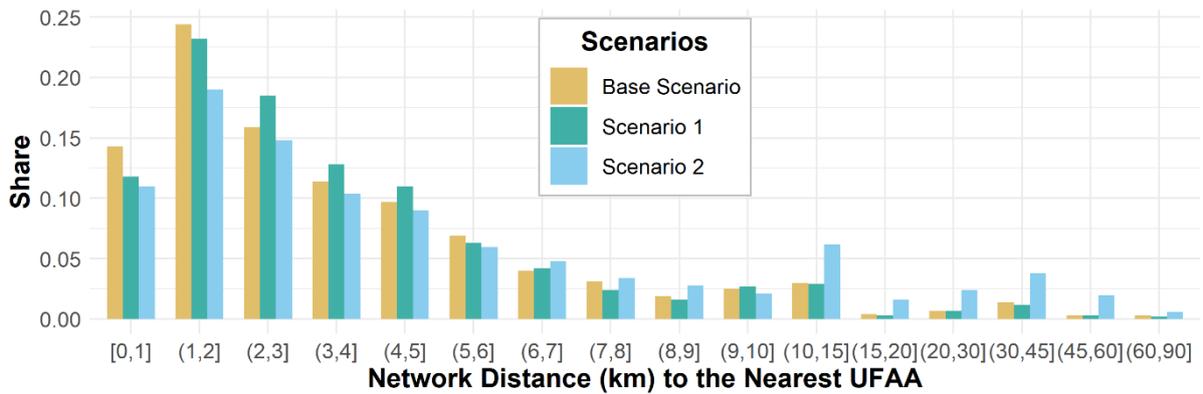

**Figure 10 Distribution of Residences' Distance to the Nearest UFAA**

**Policy Mandates to Mitigate Expansion**

Countermeasures to mitigate the urban expansion have long been discussed by the Japanese government. For example, the Location Normalization Plan of Maebashi City (*29*), the prefectural capital of Gunma, proposed to attract residents to DAA by reconstructing the decrepit buildings, embarking on redevelopment businesses, and subsidizing the residence developers and rents for students in DAA, etc. Following these, we propose two policies to attempt to mitigate the residential location expansions. The first is to grant tax exemptions for land development in DAA (hereafter Policy 1); the second is to attract the number of tertiary-sector employees from non-UFAA to UFAA (hereafter Policy 2).

The rationale behind Policy 1 is to attempt to re-balance the transport and housing cost trade-off in the residential location choice to mitigate the negative effects of increasing accessibility in suburban areas. The tax exemptions are reflected directly in the change of the land price, and will have no impact on the short-term choices, given that no long-term effect on short-term choices was assumed in this paper. Therefore, the application of Policy 1 is straightforward as only the land price attributes for the mesh cells in DAA would be modified.

A 20% decrease is proposed because it turned out to be the level that could obtain similar performance to the Base Scenario for Scenario 1. So, at least for Scenario 1, this specific value could be interpreted as the effective level to offset the expansion effect. We also consider that values higher than 20% would be less realistic to achieve, which means to attempt to offset the expansion effect in Scenario 2 was not pursued.

The rationale behind Policy 2 is that by making UFAA more attractive, the accessibilities of DAA which are generally spatially close to UFAA would increase. Also, reducing the employees in the non-UFAA would make these areas enjoy fewer accessibility, thus decreasing people's willingness to move there. One level: 30% is assumed to be increased in the UFAA from its original value. The value of 30% is proposed simply because it is the extreme value that could be imagined by imposing such a policy mandate.

Under Policy 2, the number of tertiary sector employees in UFAA is expanded by 30%. Then the number of tertiary-sector employees in non-UFAA is reduced proportionally by the weight of their respective number of tertiary-sector employees to compensate for the increases in UFAA.

Policy 2 is expected to impact the whole system in a relatively more complex way than Policy 1. First, the number of employees is one of the independent variables in the tour destination models of the short-term DAS model (*6*), hence re-running the short-term simulation is required; Second, the number of employees also impacts the independent variables of tour-based logsums in the land price model, hence the updates in land price is required.

The results of residential location models under Policies 1 and 2 are shown in **Table 5**. Short-term simulations were re-run for Policy 2, but the details are omitted for conciseness.

**TABLE 5 Simulation Results Summary under Policy Mandates.**

| Scenarios | Scenario 1 under Policy 1 | | Scenario 2 under Policy 1 | | Scenario 1 under Policy 2 | | Scenario 2 under Policy 2 | |
|---|---|---|---|---|---|---|---|---|
| **Indicators** | Value | Change rate versus Base | Value | Change rate versus Base | Value | Change rate versus Base | Value | Change rate versus Base |
| **Median network distance from residences to the closest DAA (m)** | 1,414 | +0.6% | 1,477 | +5.1% | 1,578 | +12.3% | 1,680 | +19.6% |
| **Median network distance from residences to the closest UFAA (m)** | 2,741 | +0.4% | 2,889 | +5.8% | 2,920 | +6.9% | 3,061 | +12.1% |
| **Share of Households Residing in DAA** | 38.3% | +5.8% | 36.0% | -0.6% | 31.2% | -13.8% | 29.6% | -18.2% |

The simulation results suggest that Policy 1 can significantly alleviate the residence expansion problem. For Scenario 1 where the characteristics of PAVs are assumed relatively conservative, the three indicators are found to be able to achieve a similar level to the results from Base Scenario. The share of households residing in DAA even increased by 5.8%. This effect from Policy 1 thus would satisfy, for example, the target proposed by the Location Normalization Plan of Maebashi City (*29*) that to keep the same population density in DAA between 2015 and 2040. Considerable decreases (e.g., 25.8% less in the median distance to the closest DAA compared to the case with no policy, see **Table 4**) can also be confirmed under Policy 1 in Scenario 2, though to a fewer extent than in Scenario 1. It can be summarized that providing subsidies in the land price or its equivalents could be effective to mitigate urban expansions.

Under Policy 2, the indicators surprisingly deteriorated in Scenario 1 compared to the results without no policy. By investigation, two reasons or speculations are presented. First, increasing employees in the UFAA has led to decreases in the network level of service around them, which would reduce the accessibilities and the willingness to relocate there. Second, decreasing employees in the non-UFAA would increase the chance to move there as the number of employees is a parameter with a negative coefficient in the residential location choice model (**Table 2**); this reasoning also applies to the decrease in land price in non-UFAA through the re-calculation of the land price model (**Table 1**) where the tour-based logsum variables are reduced with fewer employees.

Nevertheless, the indicators improved in Scenario 2 under Policy 2, though to a fewer extent than under Policy 1. Therefore, Policy 2 seems not as effective as Policy 1, as its effects on both transport and land use could lead to more complex changing patterns.

**CONCLUSION AND DISCUSSION**

This study made use of integrated micro-simulation models to discuss AV impacts on residential location distribution assuming the PAV prevalence in a Japanese regional area. The article focused on the long-term residential location aspect. Evaluation indicators such as median network distance to the closest urban centers and the share of households residing in urban centers are included. The connection between the short-term and long-term models is activity-based accessibility, which has not been applied much in the existing literature.

Simulation results of two AV scenarios reported moving trends away from the central areas. For example, the median network distance to Dwelling Attraction Areas increased by 7.2%, and the share of people residing in Dwelling Attraction Areas decreased by 8.3% in Scenario 1 (commuting VoT: 75%; other-purpose VoT: 85%; road capacity improvement: 1.2) compared to the Base Scenario. The shifts are considered as the re-balance of the trade-off between the housing costs and the transport costs.

Two policy mandates were tested as attempts to mitigate the moving trends. The results suggest that granting tax exemptions for land development (reducing the land prices) is found effective for both scenarios. A 20% decrease in land price would make the indicators of Scenario 1 as good as the Base Scenario. This is not the case for another policy to attract the number of tertiary-sector employees to the central areas, the results show even enlarged moving trends in Scenario 1 but improvements in Scenario 2. These findings should suggest insights for future policy making.

This work is subjected to some limitations. First, some more complex PAV behavior and the related travel behavior adaptations were not captured. For example, PAVs could be shared by another household member, potentially causing more travel and fewer accessibilities. A more sophisticated modeling tool would be required. Second, there is a large room for improvement in the land use model adopted in this study. Aspects such as job location choice, land development choice of housing or other facilities, and bidding in the household transaction would be better to incorporate to acquire more realistic forecasting in the long term. A life-cycling model to reflect the demographic pattern changes in the long term is also desirable. Nonetheless, it is necessary to acknowledge the lack of data within fine resolutions is common when conducting studies in this research context.

Furthermore, testing long-term effects on short-term choices is advised to be conducted in the future. We expect that urban expansions could be alleviated by doing so: people relocating away from the city centers would probably deteriorate the traffic level of service there hence causing a tendency to move back, though to a yet uncertain extent.

**ACKNOWLEDGMENTS**

The Gunma person trip survey used for the analysis presented in this article was provided by the Urban Planning department of Gunma Prefecture. Map data used for the analysis in this article are available from https://www.openstreetmap.org. The authors thank the two institutions for providing the data used in this study.

**AUTHOR CONTRIBUTIONS**

The authors confirm contribution to the paper as follows: study conception and design: L. Luo; data collection: L. Luo, G. Parady, K. Takami; analysis and interpretation of results: L. Luo, G. Parady; draft manuscript preparation: L. Luo. All authors reviewed the results and approved the final version of the manuscript.